\documentstyle[aps,prl,epsfig]{revtex}

\title{Classical versus Quantum Time Evolution of 
Densities at Limited Phase-Space Resolution}
 
\author{Christopher Manderfeld$^1$, Joachim Weber$^{1,2}$, and Fritz Haake$^1$}
\address{$1$ Fachbereich Physik, Universit\"at Essen, 45117 Essen, Germany\\
$2$ Department of Physics of Complex Systems, Weizmann Institute of Science, 
Rehovot 76100, Israel}

\begin{document}

\draft
\maketitle
\begin{abstract}
We study the interrelations between the classical (Frobenius-Perron)  
and the quantum (Husimi) propagator for phase-space (quasi-)probability
densities in a Hamiltonian system displaying a  mix of regular and 
chaotic behavior. We focus on common resonances of these operators
which we determine 
by blurring phase-space resolution \cite{Weber1,Weber2}.
We demonstrate that classical and quantum time evolution look alike 
if observed with a resolution  much coarser than a 
Planck cell and explain how this similarity arises for the 
propagators as well as their spectra. The indistinguishability of blurred quantum and classical evolution implies that classical resonances can conveniently be determined from quantum mechanics and in turn become effective
for decay rates of quantum correlations.  
\end{abstract}


\section{Introduction}
Classical Hamiltonian motion can appear irreversible, as is most 
familiar for many-body systems but also observed for chaos of two 
degrees of freedom. 
Effective irreversibility can show up as relaxation of certain 
correlation functions, and the decay rates involved are resonances 
of the  propagator for phase-space densities known as  the 
Frobenius-Perron operator. 
Frobenius-Perron resonances have recently been shown to play a key
role for the interrelations of classical, semiclassical and quantum 
dynamics, e.g. in a superanalytic approach to uni\-versal 
fluctuations in quantum (quasi-) energy spectra  
\cite{Altshuler,Zirnbauer,Pance}.The purpose of this paper is 
to substantiate those interrelations.

A quantum analog to the Frobenius-Perron operator 
is the propagator for Husimi functions.
Husimi functions are positive and normalized distributions
in phase space and turn into phase-space densities in the 
classical limit. 
Both classical densities and Husimi functions can be 
approximated within the same Hilbert space of phase-space functions. 
As a caveat we remark here that in the quantum case this Hilbert 
space of phase-space functions is not to be confused with 
the Hilbert space of quantum wave functions.

Even though without coupling to external reservoirs 
both the classical and the quantum 
dynamics are reversible, both do appear 
effectively irreversible when phase-space 
structures cannot be fully resolved.  
Since any measurement of phase-space variables can be conducted 
with finite precision only, restricted phase-space resolution 
is rather ubiquitous in the macroworld; in particular 
for classical chaotic dynamics, where phase-space structures extend 
over an infinite hierarchy of scales,  but also in the quantum 
case, if a Planck cell is too small to be resolved. 

By adopting finite phase-space resolution much coarser than a Planck cell 
for a prototypical dynamics we show that 
the quantum propagator becomes indistinguishable 
from the classical propagator. 
In particular, we find that resonances of the two 
propagators coincide, indicating that the same ``relaxation 
processes'' occur in classical and quantum dynamics. 
These processes and thus the resonances can be linked 
to the resolved structures in phase space \cite{Weber1,Weber2}. 
In particular, in as much as resonances are related to localized structures in phase space, our results suggest that resonances may cause some quantum eigenfunctions to assign exceptional weight to certain regions in phase or configuration space and possibly be one of the origins of scarring. 
This line of thought will be taken up in a separate paper.

If the classical phase space displays a mix of regular and chaotic structures, 
along with effective irreversible behavior caused by hyperbolicity, 
we also encounter recurrences due to elliptic islands in phase 
space; the latter manifest themselves as (almost) unimodular 
eigenvalues of the  Frobenius-Perron and Husimi propagators.

\section{Classical dynamics on the sphere}

As a prototypical Hamiltonian system with a mixed phase space we consider 
a periodically kicked angular momentum vector 
\begin{equation}
{\bf J} = (j \sin \theta \cos
\varphi, j \sin \theta \sin \varphi, j \cos \theta)
\label{eq1}
\end{equation}
of conserved length $j$, also known as the kicked top.
Such a system has one degree of freedom, and its phase space 
is the sphere, 
with the ``azimuthal'' angle $\varphi$ as the 
coordinate and the  
cosine of the ``polar'' angle $\theta$ as the conjugate momentum.
The dynamics is specified as a stroboscopic 
area-preserving map $M$ on phase space.
We choose the dynamics to consist of rotations 
$R_z(\beta_z), R_y(\beta_y)$  
about the $y-$ and 
$z-$axes by angles $\beta_y$, $\beta_z$ 
and a ``torsion'', i.e. a nonlinear
rotation $T_z(\tau) = R_z(\tau \cos \theta)$ 
about the $z-$axis which changes $\varphi$
by $\tau \cos \theta$,
\begin{equation}
M= T_z(\tau) R_z(\beta_z) R_y(\beta_y) \, .
\label{eq2}
\end{equation}
With $\beta_z$ and $\beta_y$ fixed, 
we can control the degree of 
chaoticity of the dynamics by varying $\tau$. 
While the dynamics
is integrable for $\tau=0$, with increasing $\tau$
chaoticity sets in, until for $\tau=10$ elliptic islands
have become so small that they are difficult to detect.

In the Liouville picture the time evolution of 
a phase-space density $\rho$ 
is governed by Liouville's equation,
\begin{equation}
\partial_t \rho = {\cal L} \rho = \left\{H ,\rho\right\},   
\label{eq3}
\end{equation}
where the Liouville operator ${\cal L}$, 
the Poisson bracket with the Hamiltonian $H$,
appears as the generator. 
For our rotations and torsion 
on the sphere we must separately take 
$H = \beta_y J_y, H = \beta_z J_z$, and $H = 
\frac{\tau}{2} J_z^2$.
Denoting the corresponding Liouvillians by 
${\cal L}_{R_y}$, ${\cal L}_{R_z}$, ${\cal L}_{T_z}$ 
we imagine Liouville's equation for each of them 
separately integrated over a unit time span. 
The product of the resulting three propagators yields the 
Frobenius-Perron operator 
${\cal P}=\exp({\cal L}_{T_z})\exp({\cal L}_{R_z})
\exp({\cal L}_{R_y})$.  

While a phase-space density is usually considered as 
$L^1-$integrable, it can be assumed  to belong to a 
Hilbert space of $L^2-$functions as well. 
In this Hilbert space ${\cal P}$ is represented by an
infinite dimensional unitary matrix.
Accordingly its spectrum is unimodular and, depending 
on the character of the dynamics, may consist 
of both discrete eigenvalues and continuous parts.  
The complete basis to represent the matrix in may be 
chosen as ordered with respect to phase-space 
resolution, to eventually allow for a truncation 
of ${\cal P}$ to finite size in a systematic 
manner. 

On the unit sphere the spherical harmonics 
$Y_{lm}(\theta, \varphi)$ with $l = 0, 1, 2, \dots$ 
and $|m|\le l$ form a suitable basis. Employing that represenation resonances were identified in Ref.~\cite{Weber1,Weber2}. We here briefly review those results that are of relevance for our present discussion.

Phase-space resolution is characterized by the index 
$l$: By using all $Y_{lm}$ with $0 \le l \le l_{max}$ 
phase-space structure of area $\propto 1/l_{{\rm max}}^2$ 
can be resolved. 
Our classical Liouvillians can be 
written in terms of the differential operators $\hat{L}_y, 
\hat{L}_z$ well known from quantum mechanical contexts 
in the $Y_{lm}$ representation, 
\begin{eqnarray}
{\cal L}_{R_z} & = & - {\rm i} \beta_z \hat{L}_z \,,\quad 
\hat{L}_z=-{\rm i}\frac{\partial}{\partial \varphi}\,,
\nonumber\\
{\cal L}_{R_y} & = & - {\rm i} \beta_y \hat{L}_y\,, \quad 
\hat{L}_y=-{\rm i}\left(-\cos\varphi\,
\frac{\partial}{\partial \theta}+\cot\theta\cos\varphi\,
\frac{\partial}{\partial \varphi}\right)\,,
\nonumber\\
{\cal L}_{T_z} & = & - {\rm i} \tau  \cos\theta \hat{L}_z 
\label{eq8}\,.
\end{eqnarray}
The Frobenius-Perron operator of the  $z-$rotation   
becomes the matrix
\begin{eqnarray}
\Big(\exp({\cal L}_{R_z})\Big)_{ l m, l' m'} &=& 
\int{\rm d}\theta\,\sin\theta{\rm d\varphi}\,
Y^{\ast}_{lm}(\theta,\varphi) 
\exp\left(-\beta_z\frac{\partial}{\partial\varphi}\right)
Y_{l'm'}(\theta,\varphi)\nonumber\\
&=&\delta_{l l'} \delta_{m m'} \exp(-{\rm i} m \beta_z) \, 
\label{eq9}
\end{eqnarray}
which is diagonal in both indices. 
The Frobenius-Perron matrix of the $y-$rotation is
blockdiagonal (diagonal in $l$ but not in $m$) 
and consists of the Wigner d-matrices 
well known from quantum mechanics as
\begin{equation}
\Big(\exp({\cal L}_{R_y})\Big)_{ l m, l' m'} 
= \delta_{l l'} {\rm d}^l_{m m'}(\beta_y)
\label{eq10}
\,.
\end{equation}
Finally, for the $z-$torsion the Frobenius-Perron 
matrix elements  are finite sums over products of
spherical Bessel functions $j_l(x)$ and Clebsch-Gordan 
coefficients,
\begin{equation}
\Big(\exp({\cal L}_{T_z})\Big)_{l m, l' m'} =  
\delta_{m m'} (-1)^{m} \sqrt{(2l+1)(2l'+1) } 
\sum_{l''=|l-l'|}^{l+l'}  (-{\rm i})^{l''} \; 
j_{l''}(m\tau) \; C_{0\, 0\, 0}^{l l' l''} \;
C_{-m m' 0}^{l\,\, l'\,\, l''}
\label{eq11}
\,.
\end{equation}
While both rotation matrices are diagonal in the 
indices $l, l'$, the torsion matrix ${\cal T}_{z}$
is diagonal in $m, m'$. Therefore the elements of 
the Frobenius-Perron matrix for the composite 
dynamics are the products
\begin{equation}
{\cal P}_{l m, l' m'} =\Big(\exp({\cal L}_{T_z})
\Big)_{l m, l' m} \;\Big(\exp({\cal L}_{R_z})
\Big)_{l' m, l' m}\;
\Big(\exp({\cal L}_{R_y})\Big)_{l' m, l' m'} \, ,
\label{eq12}
\end{equation}
and no infinite sum hinders their evaluation.

\section{Quantum Evolution of Husimi Functions}
In the quantum mechanical description of the kicked top 
a wave vector $|\psi_n\rangle$ is propagated by
a Floquet operator $F$ as $|\psi_{n+1}\rangle = 
F |\psi_n\rangle$ over one period in between 
two kicks.
Again $F = T_z(\tau) R_z(\beta_z) R_y(\beta_y)$ 
consists of rotation and torsion operators that 
are built of components of an angular momentum 
vector 
${\bf \hat{J}} = (\hat{J}_x, \hat{J}_y, \hat{J}_z)$ 
\cite{book} as
\begin{eqnarray}
\hat{T}_z(\tau) & = & \exp\left(-{\rm i}
\frac{\tau}{2j+1}\hat{J}_z^2\right)\, , 
\nonumber\\
\hat{R}_z(\beta_z) & = & \exp\left(-{\rm i}\beta_z 
\hat{J}_z\right) \;, 
\nonumber\\
\hat{R}_y(\beta_y) & = & \exp\left(-{\rm i}\beta_y 
\hat{J}_y\right) \;.
 \label{eq14y}
\end{eqnarray}
An obvious choice for the basis of the Hilbert 
space of wave functions are the $(2j+1)$ eigenvectors 
of $\hat{J}_z$, $\hat{J}_z |j m\rangle = m |j m\rangle$ 
with fixed $j$ and $-j \le m \le j$. 
While $\hat{T}_z(\tau)$ and $\hat{R}_z(\beta_z)$ are 
diagonal matrices in this representation, 
$\hat{R}_y$ is again given by the Wigner d-matrix 
${\rm d}^j(\beta_y)$. 
Once again we would like to emphasize that the 
Hilbert space of quantum wave functions must not 
be confused with the Hilbert space of functions 
on the classical phase space employed in Sec.II. 
(The distinction between the two Hilbert spaces 
becomes especially important when one associates, 
for integer values of $j$, the vectors $|jm\rangle$ 
with the spherical harmonics $Y_{jm}$ and the 
components of $\hat{\bf J}$ with differential 
operators.) 

As the quantum mechanical analogue to classical 
phase-space densities and the Frobenius-Perron 
operator we consider Husimi functions and their 
propagator. 
A Husimi function turns into a 
classical density in the classical limit. 
The Husimi function $Q_\rho$ of a density operator 
$\rho$ is obtained as its diagonal matrix element 
with respect to a coherent state \cite{qfunk,glauber,perelomov}.
Coherent states $|j \theta \varphi\rangle$ on the 
sphere assign to the observable $\hat{\bf J}$ a 
direction characterized by the angles $\theta$ and 
$\varphi$, but contrary to the classical description
that direction is only specified up to the minimum 
uncertainty  permitted by the angular momentum
commutation relations. The relative 
variance $( \langle\hat{\bf J}^2\rangle - 
\langle\hat{\bf J} \rangle^2 )/ j^2 = 1/j$
vanishes as the effective Planck constant $1/j$ 
goes to zero in the classical limit. 
Any coherent state $|j \theta \varphi\rangle$ can 
be obtained via a rotation $\hat{R}(\theta, 
\varphi)$ from the coherent state
$|j,m=j\rangle $ as $|j \theta \varphi\rangle = 
\hat{R}(\theta, \varphi) \, |j j\rangle$.
The Husimi function thus reads
\begin{equation}
Q_\rho(\theta, \varphi) = \langle j \theta \varphi 
| \rho | j \theta \varphi \rangle \, ;
\label{eq15}
\end{equation}
it can be expanded in terms of spherical harmonics, 
\begin{equation}
Q_\rho = \sum_{l=0}^{2j} \sum_{m=-l}^l q_{lm} Y_{lm}
\,,
\label{eq16}
\end{equation}
with the index $l$ limited to $0\le l\le 2j$ 
\cite{online}. 
The finiteness of the latter expansion will be 
commented on presently. 

As a consequence, contrary to classical densities, 
$Q_\rho$ can be represented as a finite-dimensional 
vector with the dimension set by the effective 
(inverse) Planck constant $j$ as $(2j+1)^2$. Since no
structures smaller than a Planck cell can be resolved, 
the Husimi propagator should be 
expected to be a finite matrix.

The time evolution of $Q_\rho$ is governed by von 
Neumann's equation
\begin{equation}
\partial_t Q_\rho(\theta, \varphi) 
= {\cal G} Q_\rho(\theta, \varphi) = - 
{\rm i} \langle j \theta \varphi |
\left[ \hat{H}, \rho\right] | j 
\theta \varphi \rangle \, 
\label{eq17}
\end{equation}
with $\hat{H} = \frac{\tau}{2j+1} \hat{J}_z^2, 
\hat{H} = \beta_z \hat{J}_z$ 
and $\hat{H} = \beta_y \hat{J}_y$ for torsion and 
rotations, respectively.
In the representation (\ref{eq16}) the generators 
${\cal G}$ can
be written as differential operators very similar in structure 
to the classical Liouvillians (\ref{eq8})
as (see Appendix A)
\begin{eqnarray}
{\cal G}_{T_z} & = & - {\rm i} \tau \left(\cos\theta - 
\frac{1}{2j+1} 
\frac{\partial}{\partial\theta} \sin\theta\right) 
\hat{L}_z \,,\nonumber \\
{\cal G}_{R_z} & = & - {\rm i} \beta_z 
\hat{L}_z \,,\nonumber \\
{\cal G}_{R_y} & = & - {\rm i} \beta_y 
\hat{L}_y 
\,.
\label{eq20}
\end{eqnarray}
It is seen that the quantum Husimi generators 
${\cal G}$ and the classical Liouvillians ${\cal L}$ 
are identical for rotations, while for the $z-$torsion 
a quantum correction arises in ${\cal G}_{T_z}$, 
which is formally of order $(2j+1)^{-1}$. 
This correction is responsible for all the difference 
between classical and quantum mechanics in the sequel.
We should point out here that the correction in question
ensures the finite range of  the Husimi matrix 
mentioned before, $0\le l\le 2j$, $-l\le m\le l$.
In contrast to the rotations, the torsion couples 
different angular momenta $l$. The action of the 
torsion generator on the spherical 
harmonics is calculated in Appendix B as
\begin{eqnarray}
{\cal G}_{T_z}Y_{lm}=&-&{\rm i}\frac{m\tau}{2j+1}
\left((2j-l)\sqrt{\frac{(l+m+1)(l-m+1)}{(2l+1)(2l+3)}}Y_{l+1,m}
\right. \nonumber\\ 
&+&(2j+1+l)\left.\sqrt{\frac{(l+m)(l-m)}{(2l+1)
(2l-1)}}Y_{l-1,m}\right) \,.
\label{quantgenact}
\end{eqnarray}
The generator couples spherical harmonics of 
neighbouring $l$, except for the cases $l=0$ 
and $l=2j$, where the prefactors of $Y_{l-1,m}$ 
and $Y_{l+1,m}$ vanish, respectively.

It follows from the foregoing remarks on the 
Husimi generator that the Husimi propagators 
$\exp({\cal G}_{R_z})$, $\exp({\cal G}_{R_y})$ 
for rotations  are identical to their 
Frobenius-Perron correspondents. 
However, the torsion propagator $\exp({\cal G}_{T_z})$ 
still has to be determined.
Clearly, the $(2j+1)^2$ ``Husimi functions'' 
$Q_{\left|j m_1\rangle \langle j m_2\right|}$
$=\langle j\theta\varphi |jm_1\rangle
\!\langle jm_2|j\theta\varphi\rangle$ 
are eigenfunctions of 
$\exp({\cal G}_{T_z})$ with eigenvalues
$\exp(- {\rm i} \tau (m_1^2-m_2^2) /(2j+1))$. 
Note that the $Q_{\left|j m_1\rangle 
\langle j m_2\right|}$ represents 
the skew ket-bras $\left|j m_1\rangle 
\langle j m_2\right|$ in the 
sense of the definition (\ref{eq15}).
In order to proceed to representing 
$\exp({\cal G}_{T_z})$ in the basis $\{ Y_{lm}\}$ 
we have constructed the linear transformation 
relating the spherical harmonics $Y_{lm}$ to the 
``Husimi functions'' $Q_{\left|j m_1\rangle 
\langle j m_2\right|}$ (see Appendix C).
Like its classical counterpart $\exp({\cal L}_{T_z})$ 
(see (\ref{eq11})) the Husimi propagator for $z$-torsion, 
$\exp({\cal G}_{T_z})$, turns out diagonal in $m$ but not in $l$. 
As a consequence, 
the Husimi
matrix elements are simple products of torsion and rotation 
matrix elements, 
\begin{equation}
{\cal U}_{lm,l'm'}=
\Big(\exp({\cal G}_{T_z})\Big)_{lm,l'm}
\Big(\exp({\cal G}_{R_z})\Big)_{l'm,l'm}
\Big(\exp({\cal G}_{R_y})\Big)_{l'm,l'm'}
\label{propagator}
\,.
\end{equation}

\section{Classical time evolution at finite phase-space resolution} 
\begin{figure}
\begin{center}
\epsfxsize=0.4\textwidth
\epsffile{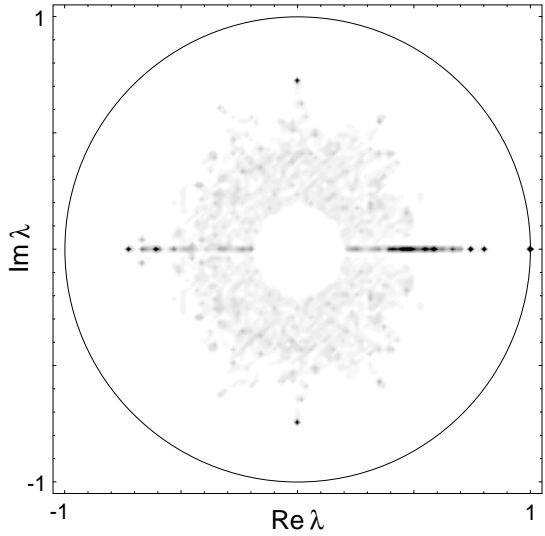}
\end{center}
\caption{For the top with $\tau = 10$, $\beta_z=\beta_y=1$  
a histogram from the eigenvalues of ${\cal P}^{(N)}$ at resolutions
$l_{max} = 20, 21 \dots 70$ in the complex 
plane has large amplitudes (black) at positions of
frozen eigenvalues. Since the mean density of eigenvalues increases drastically
near the origin, the disk with radius $1/4$ is not included in the histogram.}
\label{figure1}
\end{figure}

As has been illustrated for the kicked top in two 
previous papers \cite{Weber1,Weber2}, 
resonances and eigenvalues of the Frobenius-Perron 
operator as well as the associated phase-space 
structures can be found when time evolution is 
looked upon with finite resolution.
However, since our aim is to compare classical and 
quantum dynamics at limited phase-space resolution,
we briefly summarize the results.
In representations with respect to the 
resolution-ordered basis functions $Y_{lm}$ with 
$l=0, 1,\dots,\infty$ and integer $|m|\le l$, 
a limitation of phase-space resolution is achieved 
by a simple truncation, discarding all matrix 
elements of ${\cal P}_{l m, l' m'}$ with 
$l,l' > l_{\rm max}$.
The resulting $N\times N$ matrix ${\cal P}^{(N)}$ 
of dimension $N = (l_{\rm max}+1)^2$ is a nonunitary 
approximation to the unitary ${\cal P}$ with a purely
discrete spectrum bounded in modulus by unity. 
An investigation of the $N-$dependence of the 
eigenvalues of ${\cal P}^{(N)}$ reveals that with 
increasing resolution some eigenvalues persist 
in their positions, either close to or well inside 
the unit circle. Of these ``frozen'' eigenvalues the 
almost unimodular ones turn into unimodular 
eigenvalues of  the unitary ${\cal P}$ as 
$N\to\infty$; the subunimodular eigenvalues reflect 
resonances of ${\cal P}$ in a higher Riemannian sheet.
The different nature of ``frozen'' unimodular and 
subunimodular eigenvalues becomes more evident 
through an investigation of their eigenfunctions. 
For unimodular eigenvalues the eigenfunctions are
sharply localized on islands of regular motion 
surrounding  elliptic periodic orbits, while
for subunimodular eigenvalues eigenfunctions are 
localized around unstable manifolds of hyperbolic 
periodic orbits.
The condition for freezing of an eigenvalue 
evidently is that at least the largest associated 
phase-space structures are resolved. 
For the kicked top with $\tau = 10$, $\beta_z = 1$, 
and $\beta_y = 1$ we propose to illustrate the freezing of 
eigenvalues. 
As no elliptic structures can be resolved even at
the highest employed resolution, $l_{max} = 70$, 
all frozen eigenvalues except for the one 
at unity (pertaining to the stationary constant eigenfunction
$Y_{00}$) lie well inside the unit circle. A gray-scale 
histogram of all eigenvalues with moduli greater 
than $1/4$ of all matrices ${\cal P}^{(N)}$ 
with $l_{max} = 20, 21 \dots, 70$ in the complex plane 
is shown in figure \ref{figure1}. Eigenvalues of 
smaller modulus have been rejected, since
they have not settled yet and would spoil the histogram 
due to their large density near the origin.
Dark areas in the histogram indicate large amplitudes
and thus the positions of ``frozen'' eigenvalues. 
Several nonunimodular ``frozen'' eigenvalues are 
clearly  visible. Their precise positions at 
resolutions $l_{\rm max} = 30, 40, 50, 60,$ and $68$ 
are given in table \ref{table1}.

\section{Quantum time evolution at blurred phase-space resolution}
The smallest scale in phase space resolvable 
in Husimi functions 
is set by Planck's constant. Recalling that for 
the kicked top the role of the inverse Planck constant 
is played by $j$, in the Hilbert space spanned by the $Y_{lm}$ 
the Husimi propagator can be  represented 
as a finite $N\times N$ matrix of dimension 
$N=(2j+1)^2$. This is because with respect to 
the basis  functions $Y_{lm}$ all matrix elements 
of ${\cal U}$ involving indices $l > 2j$ vanish.  
The $(2j+1)^2$ discrete eigenvalues of ${\cal U}$ 
are of unit modulus due to the unitarity of 
Schr{\"o}dinger's time evolution which also implies 
conservation of 
probability in phase space. With the eigenvalue 
problem for the Floquet operator $F|\phi_k\rangle = 
\exp(-{\rm i}\phi_k) |\phi_k\rangle$
solved, the eigenphases  of the full ${\cal U}$ 
are the quasienergy differences $(\phi_k - \phi_l)$ and the 
associated eigenfunctions follow as $\langle j 
\theta \varphi | \phi_k\rangle\langle \phi_l | j 
\theta \varphi \rangle$.

We truncate 
the propagator ${\cal U}$ in the same manner 
as the Frobenius-Perron matrix, with 
the resolution parameter $l_{max}$ taking a value 
$l_{max}<2j$. Now the dependence  of the eigenvalues 
on the truncation parameter $l_{max}$ can be studied. 
We are particularly interested in semiclassical 
propagators, i.e. large values of $j$. Introducing 
the ratio $\kappa = l_{max}/(2j)$, 
$\kappa = 1$ represents complete resolution and $\kappa \ll 1$ 
the limit with Planck cells far from resolved. 

\begin{figure}
\begin{center}
\epsfxsize=0.7\textwidth
\epsffile{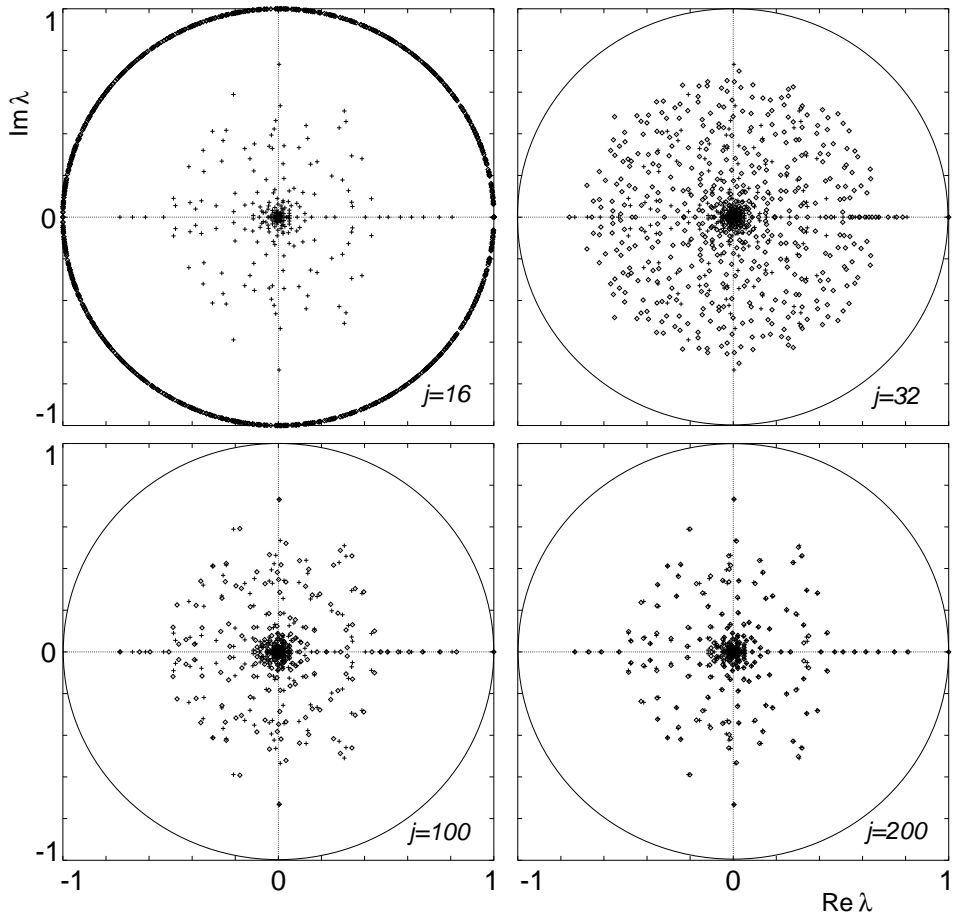}
\end{center}
\caption{Eigenvalues of the truncated Husimi $(\diamond)$ and 
Frobenius-Perron $(+)$ operators ($\tau = 10, \beta_z = 1,
\beta_y = 1$) superimposed at $l_{max}=32$
with the quantum number $j$ taking the values $16$ (full 
resolution), $32, 100, 200$. While for $j=16$ the Husimi 
spectrum is unimodular, for $j=200$ classical and quantum 
eigenvalues are in good agreement.
 }
\label{figure4}
\end{figure}

Starting from $\kappa=1$ and slowly decreasing 
$\kappa$, the eigenvalues of ${\cal U}$ feel the 
truncation as a small perturbation at first that 
slightly shifts them 
inside the unit circle. With a further decrease 
of $\kappa$, eigenvalues are spread over the entire 
unit disk. When $\kappa$ finally becomes small, 
the eigenvalues of the truncated propagator matrix 
coincide with the eigenvalues of the corresponding 
Frobenius-Perron matrix ${\cal P}^{(N)}$ of the 
same dimension. This is illustrated in figure 
\ref{figure4}, where the Husimi eigenvalues 
of the kicked top with $\tau = 10$, 
$\beta_z = \beta_y = 1$ for $\kappa = 1, 0.5, 0.16$  
and $0.08$ are compared to the eigenvalues of the 
Frobenius-Perron matrix at $l_{max}=32$. For the 
sake of the comparison we kept $l_{max} = 32$ fixed 
also for the quantum propagator and changed 
$\kappa$ by choosing different quantum numbers 
$j = 16, 32, 100, 200$.  As the most striking aspect 
of the spectral coincidence at small $\kappa$, 
resonances as well as eigenvalues of the full 
Frobenius-Perron operator can be obtained from 
quantum dynamics by looking for frozen 
eigenvalues of the truncated Husimi propagator at 
different values $l_{max}$ in the limit
$\kappa \ll 1$.  Furthermore, the coincidence of 
eigenvalues implies that also the associated 
eigenfunctions are in agreement.
In particular, the eigenfunctions of the truncated 
Husimi propagator display the same strong scarring 
on unstable manifolds and elliptic islands in 
phase space as their classical  counterparts.

\begin{figure}
\begin{center}
\epsfxsize=0.5\textwidth
\epsffile{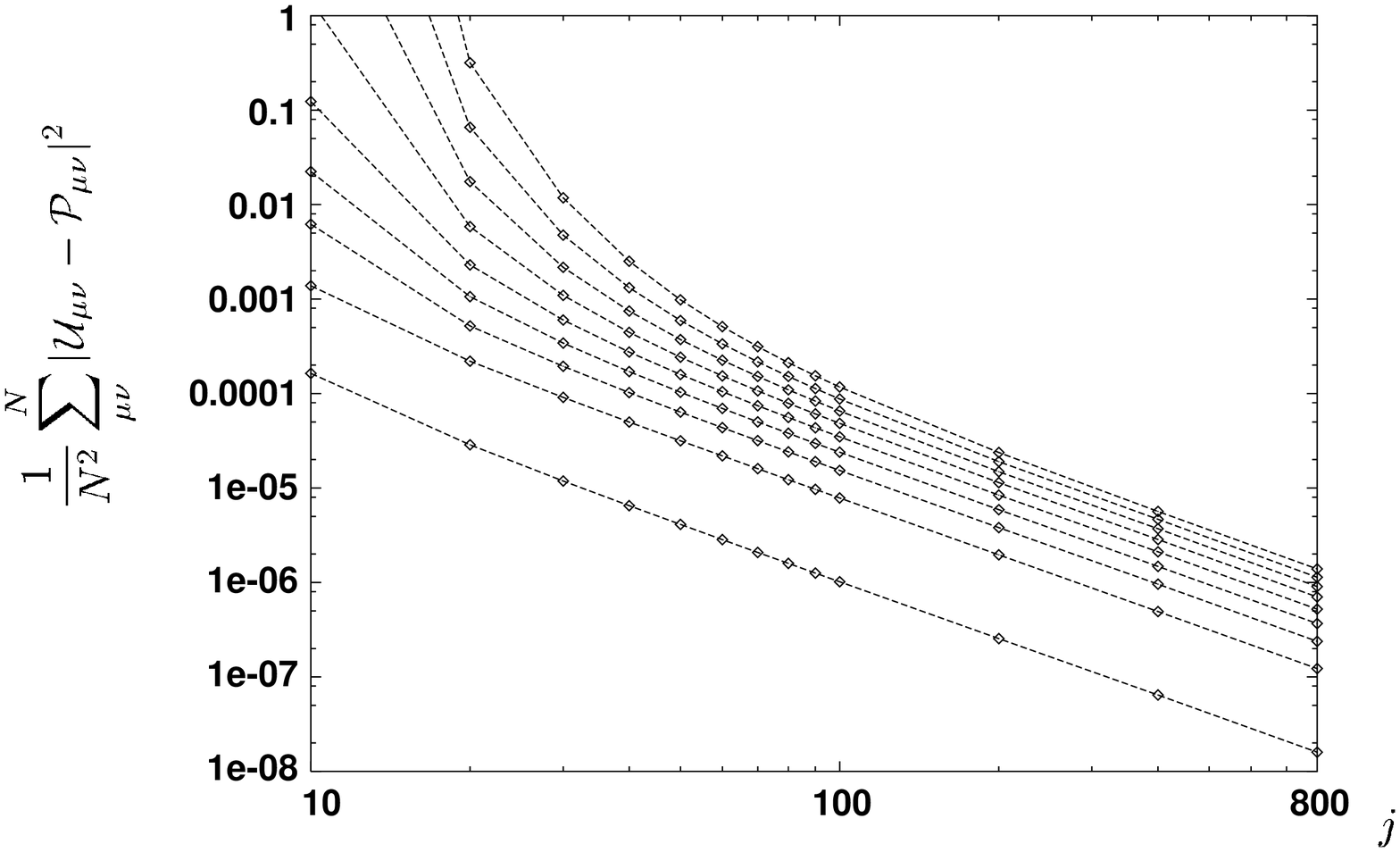}
\end{center}
\caption{Mean squared deviation between Husimi and 
Frobenius-Perron matrix elements of the top ($\tau = 10, 
\beta_z = 1, \beta_y = 1$) with $l_{max} = 4, 6, 8, 10, 12, 
14, 16, 18, 20$ as a function of $j$ in a double-logarithmic
plot. The numerical data suggest a power-law decay of the 
mean squared deviation as $j\to \infty$.
}
\label{figure5}
\end{figure}

The most drastic difference between the two propagator 
matrices lies, of course, in the finiteness of the quantum 
propagator. Moreover, the matrix elements with $l$ near the quantum 
cutoff $2j$ differ appreciably: It is illustrated in 
Appendix B that the quantum correction of the
torsion generator in the basis of spherical harmonics
is of the order $l/(2j+1)$ relative
to the classical part. 
On the other hand, on large phase-space scales, i.e. for $l\ll 2j$, 
the Frobenius-Perron and Husimi propagators become 
indistinguishable as $\kappa\to 0$. 
An expansion of  the Husimi propagator in powers of $(2j+1)^{-1}$
yields
\begin{equation}
\Big(\exp({\cal G})\Big)_{lm,l'm'}=
\Big(\exp({\cal L})\Big)_{lm,l'm'}
+\frac{1}{2j+1}\Big(\sum\dots\Big)_{lm,l'm'}
\;.\end{equation}
Assuming $l_{max}$ fixed, for $j\to \infty$ the quantum 
correction of a matrix element decays as $1/(2j+1)$. 
Accordingly, in this case we expect the mean squared deviation 
between the matrix elements of the two propagators to 
asymptotically vanish as  $1/(2j+1)^2$.
Figure \ref{figure5} confirms that expectation. 

\section{Conclusion}
In conclusion, for a prototypical dynamical system 
we have formulated time evolution in terms of phase-space 
distributions, both classically and quantum mechanically. 
The comparison of the two respective propagators at 
limited phase-space resolution unveils the classical 
character of the quantum dynamics on large phase-space scales. 
Only when quantum coherences on the scale of Planck's 
constant are resolved, the peculiarities of 
quantum mechanics arise. 
Due to this fact 
Frobenius-Perron resonances can be identified 
from the quantum propagator. 
Moreover, quantum propagation 
is described by a finite matrix and therefore more easily treated 
than the classical counterpart which is an infinite matrix. 
Our results on classical and quantum signatures of resonances 
might become helpful in explaining 
phase-space localization (scars) 
of quantum eigenfunctions. 

We gratefully acknowledge support by the Sonderforschungsbereich 
`Unordnung und gro{\ss}e Fluktuationen' of the Deutsche 
Forschungsgemeinschaft and support of the Minerva foundation. 

\section{Appendix}
\subsection{Husimi generators for rotation and torsion}

A normalized coherent state can be written as 
\cite{qfunk,glauber,perelomov,online} 
\begin{equation}
|j\theta\varphi\rangle=(1+\alpha\alpha^\ast)^{-j}{\rm e}^{\alpha \hat{J}_-}
|jj\rangle=(1+\alpha\alpha^\ast)^{-j}\sum_{m=-j}^j\alpha^{j-m}
\sqrt{2j\choose j-m}|jm\rangle
\label{coherent}\,,
\end{equation}
where $\alpha$ is the complex parameter  
$\alpha=\tan\frac{\theta}{2}\,{\rm e}^{{\rm i}\varphi}$ and 
$\hat{J}_\pm =\hat{J}_x\pm {\rm i}\hat{J}_y$ are the 
familiar ladder operators.
In this manner a Husimi function (\ref{eq15}) can be expressed as 
a function of $\alpha$ and its complex conjugate $\alpha^\ast$,
\begin{equation}
Q_{\rho}(\alpha,\alpha^\ast)=(1+\alpha\alpha^\ast)^{-2j}
\langle jj|{\rm e}^{\alpha^\ast \hat{J}_+}\,\rho
\,{\rm e}^{\alpha \hat{J}_-}|jj\rangle
\label{hus}\,.
\end{equation}
In order to derive the generators defined by von Neumann's 
equation (\ref{eq17}) we start with the following two identities
\begin{eqnarray}
\langle jj|{\rm e}^{\alpha^\ast \hat{J}_+}\,\rho 
\hat{J}_z\,{\rm e}^{\alpha \hat{J}_-}|jj\rangle
&=&\langle jj|{\rm e}^{\alpha^\ast \hat{J}_+}\,\rho\,
{\rm e}^{\alpha \hat{J}_-}(\hat{J}_z-\alpha\hat{J}_-)|jj\rangle
\nonumber\\
&=&\left(j-\alpha\frac{\partial}{\partial \alpha}\right)
\langle jj|{\rm e}^{\alpha^\ast \hat{J}_+}\,\rho
\,{\rm e}^{\alpha \hat{J}_-}|jj\rangle
\,,\nonumber\\
\langle jj|{\rm e}^{\alpha^\ast \hat{J}_+}\,\hat{J}_z\rho 
\,{\rm e}^{\alpha \hat{J}_-}|jj\rangle
&=&\langle jj|(\hat{J}_z-\alpha^\ast\hat{J}_+)
{\rm e}^{\alpha^\ast \hat{J}_+}\,\rho\,
{\rm e}^{\alpha \hat{J}_-}|jj\rangle
\nonumber\\
&=&\left(j-\alpha^\ast\frac{\partial}{\partial \alpha^\ast}\right)
\langle jj|{\rm e}^{\alpha^\ast \hat{J}_+}\,\rho
\,{\rm e}^{\alpha \hat{J}_-}|jj\rangle
\label{relation2}
\,.
\end{eqnarray}
Thus, the generator for a rotation about the $z$-axis becomes
\begin{eqnarray}
\frac{{\rm i}}{\beta_z}{\cal G}_{R_z}Q_{\rho}&=&
(1+\alpha\alpha^\ast)^{-2j}
\langle jj|{\rm e}^{\alpha^\ast \hat{J}_+}\left[\hat{J}_z,\rho\right]
{\rm e}^{\alpha \hat{J}_-}|jj\rangle
\nonumber\\
&=&(1+\alpha\alpha^\ast)^{-2j}
\left[\left(j-\alpha^\ast\frac{\partial}{\partial \alpha^\ast}\right)
-\left(j-\alpha\frac{\partial}{\partial \alpha}\right)\right]
\langle jj|{\rm e}^{\alpha^\ast \hat{J}_+}\,\rho
\,{\rm e}^{\alpha \hat{J}_-}|jj\rangle
\,.
\end{eqnarray}    
It is easy to see that the differential operator commutes with 
the prefactor $(1+\alpha\alpha^\ast)^{-2j}$ such that
\begin{equation}
\frac{{\rm i}}{\beta_z}{\cal G}_{R_z}Q_{\rho}=
\left(\alpha\frac{\partial}{\partial\alpha}-
\alpha^\ast\frac{\partial}{\partial\alpha^\ast}\right)Q_{\rho}
\label{commutes}\,.
\end{equation} 
Replacing the complex variables $\alpha$, $\alpha^\ast$ by the 
angular coordinates $\theta$, $\varphi$ we finally get
\begin{equation}
\alpha\frac{\partial}{\partial \alpha}
- \alpha^\ast\frac{\partial}{\partial \alpha^\ast}
=-{\rm i}\frac{\partial}{\partial\varphi}
\,,
\end{equation}
whereupon the generator becomes 
\begin{equation}
\frac{{\rm i}}{\beta_z}{\cal G}_{R_z}
=-{\rm i}\frac{\partial}{\partial \varphi}=\hat{L}_z
\,.
\end{equation}  

Some more effort is required for the torsion generator, 
\begin{equation}
{\rm i}\frac{2j+1}{\tau}{\cal G}_{T_z}Q_{\rho}=
(1+\alpha\alpha^\ast)^{-2j}
\langle jj|{\rm e}^{\alpha^\ast \hat{J}_+}\left[\hat{J}_z^2,\rho\right]
{\rm e}^{\alpha \hat{J}_-}|jj\rangle
\,.
\end{equation} 
After using (\ref{relation2}) twice one obtains
\begin{eqnarray}
{\rm i}\frac{2j+1}{\tau}{\cal G}_{T_z}Q_{\rho}
&=&(1\!+\!\alpha\alpha^\ast)^{-2j}
\left[\left(j-\alpha^\ast\frac{\partial}{\partial \alpha^\ast}\right)^2
\!-\!\left(j-\alpha\frac{\partial}{\partial \alpha}\right)^2\right]
\langle jj|{\rm e}^{\alpha^\ast \hat{J}_+}\,\rho
\,{\rm e}^{\alpha \hat{J}_-}|jj\rangle
\nonumber\\
&=&(1\!+\!\alpha\alpha^\ast)^{-2j}
\left[\left(\alpha^\ast\frac{\partial}{\partial\alpha^\ast}\right)^2
\!-\!\left(\alpha\frac{\partial}{\partial\alpha}\right)^2
\!+\!2j\alpha\frac{\partial}{\partial\alpha}
\!-\!2j\alpha^\ast\frac{\partial}{\partial\alpha^\ast}\right]
\langle jj|{\rm e}^{\alpha^\ast \hat{J}_+}\,\rho
\,{\rm e}^{\alpha \hat{J}_-}|jj\rangle
\nonumber\\
&=&(1\!+\!\alpha\alpha^\ast)^{-2j}
\left(\alpha^\ast\frac{\partial}{\partial\alpha^\ast}+
\alpha\frac{\partial}{\partial \alpha}-2j\right)
\!\left(\alpha^\ast\frac{\partial}{\partial\alpha^\ast}
-\alpha\frac{\partial}{\partial\alpha}\right)
\langle jj|{\rm e}^{\alpha^\ast \hat{J}_+}\,\rho
\,{\rm e}^{\alpha \hat{J}_-}|jj\rangle
\,.
\end{eqnarray}
The commutation of the left differential operator with the prefactor 
generates a further term, while the right one commutes as in (\ref{commutes}),
\begin{equation}
{\rm i}\frac{2j+1}{\tau}{\cal G}_{T_z}Q_{\rho}
=\left(2j\frac{2\alpha\alpha^\ast}{1+\alpha\alpha^\ast}-2j
+\alpha\frac{\partial}{\partial \alpha}
+\alpha^\ast\frac{\partial}{\partial\alpha^\ast}\right)
\left(\alpha^\ast\frac{\partial}{\partial\alpha^\ast}-
\alpha\frac{\partial}{\partial \alpha}\right)Q_{\rho}
\,.
\end{equation}
Again we replace the complex coordinates by the spherical
coordinates,
\begin{eqnarray}
\alpha\frac{\partial}{\partial \alpha}
+\alpha^\ast\frac{\partial}{\partial\alpha^\ast}
&=&\sin\theta\frac{\partial}{\partial\theta}\,,\\
1-\frac{2\alpha\alpha^\ast}{1+\alpha\alpha^\ast}
=\frac{1-\alpha\alpha^\ast}{1+\alpha\alpha^\ast}
&=&\cos\theta
\,,
\end{eqnarray}
and the generator finally becomes 
\begin{equation}
{\cal G}_{T_z}= -\tau \left(\cos\theta-\frac{1}{2j+1}
\frac{\partial}{\partial\theta}\,\sin\theta\right)
\frac{\partial}{\partial\varphi}
\,.
\label{quantumgenerator}
\end{equation}

\subsection{The torsion generator matrix} 
\label{appb}
We here calculate the action of the classical and quantum
torsion generators on spherical harmonics. 
Spherical harmonics are separable into $\varphi$- and 
$\theta$-dependent parts as
\begin{equation}
Y_{lm}(\theta,\varphi)=\sqrt{\frac{2l+1}{4\pi}\frac{(l-m)!}{(l+m)!}}
P_l^m(\cos\theta){\rm e}^{{\rm i}m\varphi}
\label{sphere}\,,
\end{equation}
where $P_l^m(\cos\theta)$ are the associated Legendre polynomials \cite{abram}. 
At first we focus on the more complicated action of 
the generators ${\cal L}_{T_z}$ and ${\cal G}_{T_z}$ on the $P_l^m(\cos\theta)$, 
i.e. for the classical propagator 
\begin{equation}
\cos\theta \, P_l^m(\cos\theta)  =  z P_l^m(z)\,,
\end{equation}
and for the quantum correction (cf. (\ref{quantumgenerator}))
\begin{equation}
-\frac{1}{2j+1}\frac{\partial}{\partial\theta}\,\sin\theta \, P_l^m(\cos\theta) = 
- \frac{1}{2j+1}\left(z - (1-z^2)\frac{\rm d}{{\rm d}z}\right)P_l^m(z) \,.
\end{equation}
By using the following recursion formulae for the $P_l^m(z)$ \cite{abram}, 
\begin{eqnarray}
(1-z^2)\frac{\rm d}{{\rm d}z}P_l^m(z)=
-lzP_{l}^m(z)+(l+m)P_{l-1}^m(z)\,,\\
(2l+1)zP_l^m(z)=(l-m+1)P_{l+1}^m+(l+m)P_{l-1}^m(z)
\, ,
\end{eqnarray}
one arrives at
\begin{equation}
z P_l^m(z) = \frac{l-m+1}{2l+1} P_{l+1}^m(z) + 
\frac{l+m}{2l+1} P_{l-1}^m(z) 
\end{equation}
for the classical generator and 
\begin{equation}
- \frac{1}{2j+1}\left(z - (1-z^2)\frac{\rm d}{{\rm d}z}\right)P_l^m(z) = 
-\frac{1}{(2j+1) (2l+1)} \Big((l+1)(l-m+1) P_{l+1}^m(z) - 
l (l+m) P_{l-1}^m(z) \Big)
\label{classgenact}
\end{equation}
for the quantum correction term.
The associated Legendre polynomials can now be replaced by the
spherical harmonics (\ref{sphere}). Taking into account  
$\hat{L}_z Y_{lm}= m Y_{lm}$, we finally see the classical torsion generator 
to act on the spherical harmonics as
\begin{equation}
{\rm i} \frac{{\cal L}_{T_z}}{\tau m} Y_{lm} = 
\sqrt{\frac{(l-m+1) (l+m+1)}{(2l+1) (2l+3)}} Y_{(l+1) m} +
\sqrt{\frac{(l-m)(l+m)}{(2l+1) (2l-1)}} Y_{(l-1)m} \, ,
\end{equation}
and the quantum correction as 
\begin{equation}
{\rm i} \frac{{\cal G}_{T_z} - {\cal L}_{T_z}}{\tau m} Y_{l m}  = 
-\frac{l+1}{2j+1} \sqrt{\frac{(l-m+1)(l+m+1)}{(2l+3)(2l+1)}} Y_{(l+1) m} +
\frac{l}{2j+1} \sqrt{\frac{(l-m)(l+m)}{(2l-1)(2l+1)}} Y_{(l-1) m} \,.
\label{quantgencorr}
\end{equation}
The expression for ${\cal G}_{T_z} Y_{lm}$ is already given in (\ref{quantgenact}).
The matrix elements of ${\cal L}_{T_z}$ and ${\cal G}_{T_z}$ in the basis of spherical harmonics
can easily be read off from (\ref{classgenact}) and (\ref{quantgenact}), respectively.
It is noteworthy that relative to the classical matrix element the quantum correction, to 
be read off from (\ref{quantgencorr}),
is of order $l/(2j+1)$. 
Therefore matrix elements of the quantum generator with $l\sim 2j$ differ 
significantly from their classical counterparts; 
most importantly, the quantum correction manifests the finiteness of the 
expansion of $Q$ at $l=2j$, in accordance with the remarks after (\ref{quantgenact}).

\subsection{Husimi torsion propagator}
\label{torsmat}
In order to obtain the propagator matrix (\ref{propagator}) the torsion 
propagator $\exp({\cal G}_{T_z})$ is to be written in the basis of spherical harmonics. 
Therefore we start from the eigenrepresentation of the propagator and apply
the respective linear transformation.
The basis of right-hand eigenfunctions consists of the  $(2j+1)^2$ ``Husimi 
functions'' 
\begin{equation}
Q_{|jm_1\rangle\langle jm_2|} = \langle j \theta \varphi |j m_1\rangle 
\langle j m_2 | j \theta \varphi \rangle \, , 
\end{equation}
with eigenvalues $\exp(-{\rm i}\tau(m_1^2-m_2^2)/(2j+1))$. 

The left-hand eigenfunctions lie in the function space dual to the space 
of Husimi functions. Functions from this space are the so-called $P$ functions 
\cite{qfunk,glauber,perelomov,online}. For an operator $\rho$ the $P$ function
$P_{\rho}(\theta,\varphi)$ is defined as the weight of 
the diagonal mixture with respect to 
coherent states, 
\begin{equation}
\rho = \frac{2j+1}{4 \pi} \int_0^{\pi}{\rm d}\theta\,
\sin\theta \,\int_0^{2\pi}{\rm d}\varphi\, 
P_{\rho}(\theta,\varphi) \,
|j\theta\varphi\rangle\langle j\theta\varphi| \, .
\end{equation}
In particular, the functions $P_{|jm_1\rangle\langle jm_2|}$  
are the left-hand eigenfunctions of $\exp({\cal G}_{T_z})$ and form a basis
of this dual space that is biorthonormal to the basis 
$Q_{|jm_1\rangle\langle jm_2|}$.
Therefore, the eigenrepresentation of the torsion propagator is given by \cite{online}
\begin{equation}
\exp({\cal G}_{T_z})=\sum_{m_1,m_2=-j}^j
|Q_{|jm_1\rangle\langle jm_2|}\rangle\!\rangle
\exp\left(-{\rm i}\tau\frac{m_1^2-m_2^2}{2j+1}\right)
\langle\!\langle P_{|jm_1\rangle\langle jm_2|}|
\, ,
\end{equation}
wherein the double brackets refer to the scalar product of the classical 
Hilbert space.

The coefficients $q_{lm}(m_1,m_2)$ of the linear transformation connecting 
Husimi functions and spherical harmonics  and the coefficients 
$p_{lm}(m_1,m_2)$ of the inverse transformation can be  obtained as 
\cite{qfunk,perelomov,online}
\begin{eqnarray}
q_{lm}(m_1,m_2) & = & 
\langle\!\langle Y_{lm}|Q_{|jm_1\rangle\langle jm_2|}\rangle\!\rangle
=\int{\rm d}\theta\,{\rm d}\varphi\,
\sin\theta \, Y^*_{lm} Q_{|jm_1\rangle\langle jm_2|}(\theta,\varphi) \,, \\
p_{lm}(m_1,m_2) & = & 
\langle\!\langle Y_{lm}|P_{|jm_1\rangle\langle jm_2|}\rangle\!\rangle
\int{\rm d}\theta\,{\rm d}\varphi\,
\sin\theta \, Y^*_{lm} P_{|jm_1\rangle\langle jm_2|}(\theta,\varphi) \, .
\end{eqnarray}
While the first integral yields the real coefficients 
(products of Clebsch-Gordan coefficients, incidentally \cite{qfunk,perelomov})
\begin{eqnarray}
q_{lm}(m_1,m_2)&=& \delta_{m,m_1\!-\!m_2} \sqrt{4\pi(2l+1)\frac{(l-m)!}{(l+m)!}}
\sqrt{{2j\choose j\!-\!m_1}{2j\choose j\!-\!m_2}}
\nonumber\\
&&\times \sum_{k=0}^{l-m}
\frac{(-1)^{k+m}(l+m+k)!(j+m_1)!(j+k-m_2)!}
{(l-m-k)!k!(m+k)!(2j+k+m+1)!}
\,, 
\label{qexplizit}
\end{eqnarray}
the coefficients $p_{l m}(m_1,m_2)$ are proportional to the $q_{l m}(m_1,m_2)$,
\begin{equation}
p_{l m}(m_1,m_2) = \frac{\sqrt{4\pi(2l+1)}}{(2j+1)q_{l0, j j}}
q_{lm}(m_1,m_2)
\,.
\end{equation}

Actually, the above formulae are not particularly useful 
for numerical purposes as they 
involve an alternating sum of very large numbers. 
Easier to employ are 
recurrence formulae for the coefficients. 
One formula we used reads
\begin{equation}
q_{l(m-1)}(m_1,m_2) = \frac{1}{w(l,-m)} \Big(w(j,m_1) q_{lm}(m_1+1,m_2) 
- w(j,-m_2) q_{lm}(m_1,m_2-1)\Big)
\,,
\end{equation}
wherein  $w(l,m) = \sqrt{l(l+1)-m(m+1)}$. The recursion starts with
\begin{equation}
q_{ll}(m_1,m_2) = \delta_{l, m_1\!-\!m_2}  \sqrt{4 \pi (2l+1)} \frac{(2j)! 
\sqrt{(2l)!}}{(2j+l+1)!\, l!} \sqrt{\frac{(j+m_1)!
(j-m_2)!}{(j-m_1)! (j+m_2)!}} \, .
\end{equation}
Coefficients with negative values of $m$ can be obtained from ones 
with positive $m$ through the symmetry relation
\begin{equation}
q_{l(-m)}(m_1,m_2) = (-1)^{m_1-m_2} q_{lm}(m_1,m_2) \, .
\end{equation}
Finally, the propagator matrix in the basis of spherical harmonics is obtained as
\begin{eqnarray}
\Big(\exp({\cal G}_{T_z})\Big)_{lm,l'm'}=\delta_{m,m'}
\sum_{m_1={\rm max}(-j+m,-j)}^{{\rm min}(j,j+m)}
q_{lm}(m_1,m_1-m) \,\, p_{l'm}(m_1,m_1-m)
\exp\left(-{\rm i}\tau\frac{2mm_1-m^2}{2j+1}\right)
\,.
\end{eqnarray}


\begin{table}
\[
\begin{array}{|r|r|r|r|r|} \hline
l_{max} = 30 & l_{max} = 40 & l_{max} = 50 & l_{max} = 60 & l_{max} = 68 \\ \hline\hline
0.8018 & 0.8116 & 0.8205 & 0.8103 & 0.8074 \\ \hline
0.7513  & 0.7457 & 0.7547 & 0.7470 & 0.7455 \\ \hline
\begin{array}{r} -0.0041\\ \pm {\rm i}\; 0.7225\end{array} & 
\begin{array}{r} -0.0063\\ \pm {\rm i}\; 0.7431\end{array} &
\begin{array}{r} -0.0123\\ \pm {\rm i}\; 0.7515\end{array} &
\begin{array}{r} -0.0079\\ \pm {\rm i}\; 0.7517\end{array} &
\begin{array}{r} -0.0027\\ \pm {\rm i}\; 0.7435\end{array} \\ \hline
-0.7391 & -0.7432 & -0.7475 & -0.7510 & -0.7427 \\ \hline
-0.6639 & -0.6766 & -0.6746 & -0.6869 & -0.6955 \\ \hline
0.6470 &  0.6628 &  0.6777 &  0.6597 &  0.6727 \\ \hline
-0.6218 & -0.6443 & -0.6188 & -0.6377 & -0.6336 \\ \hline
0.5462 &  0.5968 &  0.5847 &  0.5889 &  0.5786 \\ \hline
-0.5141 & -0.5406 & -0.5363 & -0.5481 & -0.5448 \\ \hline
\end{array}
\]
\caption{Frozen nonunimodular eigenvalues of ${\cal P}^{(N)}$ with 
$\tau = 10$, $\beta_z=\beta_y=1$ 
at the resolutions $l_{max} = 30, 40, 50, 60, 68$.}
\label{table1}
\end{table}

\begin{thebibliography}{99}

\bibitem{Weber1}
J. Weber, F. Haake, and P. \v{S}eba, Phys. Rev. Lett. {\bf 85}, 3620 (2000)
\bibitem{Weber2}
J. Weber, F. Haake, P.~A. Braun, C. Manderfeld, and P. \v{S}eba, 
J. Phys. A {\bf 34}, 7195 (2001)
\bibitem{Altshuler} A.~V. Andreev and B.~L. Altshuler, Phys. Rev. Lett. {\bf
75}, 902 (1995);
O. Agam, B.~L. Altshuler, and A.~V. Andreev, Phys. Rev. Lett. {\bf 75}, 4389
(1995);
A.~V. Andreev,O. Agam, B.~D. Simons, and B.~L. Altshuler, Phys. Rev. Lett. {\bf 76}, 1 (1996);
A.~V. Andreev, B.~D. Simons, O. Agam, and B.~L. Altshuler,  Nuclear Physics B, {\bf 482}, 536 (1996).
\bibitem{Zirnbauer}
M.~R. Zirnbauer in: I.V. Lerner,
J.P. Keating, and D.E. Khmelnitskii (eds.), {\em Supersymmetry and Trace
Formulae:  Chaos and Disorder}
(Kluwer Academic, New York, 1999)
\bibitem{Pance} 
K.Pance, W. Lu, S. Sridhar, Phys. Rev. Lett. {\bf 85}, 2737 (2000)
\bibitem{book} F. Haake, {\em Quantum Signatures of Chaos} 
(Springer, Berlin, 2001)
\bibitem{qfunk} F. T. Arecchi, E. Courtens, 
R. Gilmore, H. Thomas, Phys. Rev. A {\bf 6}, 2211 (1972) 
\bibitem{glauber} R. Glauber and F. Haake, Phys. Rev. A {\bf 13}, 357 (1976)
\bibitem{perelomov} A. M. Peremolov, {\em Generalized Coherent States and Their Applications} 
(Springer, New York, 1986)
\bibitem{online} C. Manderfeld,{\em Coherent State Representation of the 
$SU(2)$ Group}, http://www.theo-phys.uni-essen.de/tp/u/chris/
\bibitem{abram} M. Abramowitz and I. A. Stegun, {\em Handbook of Mathematical 
Functions} (Dover, New York, 1972) 
\end{thebibliography}
\end{document}